\documentclass[sigplan,10pt]{acmart}
\usepackage[utf8]{inputenc}
\usepackage{url}
\usepackage{hyperref}
\usepackage{listings}
\usepackage{GoMono}

\lstdefinelanguage{Golang}%
  {morekeywords=[1]{package,import,func,type,struct,return,defer,panic,%
     recover,select,var,const,iota,},%
   morekeywords=[2]{string,uint,uint8,uint16,uint32,uint64,int,int8,int16,%
     int32,int64,bool,float32,float64,complex64,complex128,byte,rune,uintptr,%
     error,interface},%
   morekeywords=[3]{map,slice,make,new,nil,len,cap,copy,close,true,false,%
     delete,append,real,imag,complex,chan,},%
   morekeywords=[4]{for,break,continue,range,go,goto,switch,case,fallthrough,if,%
     else,default,},%
   morekeywords=[5]{Println,Printf,Error,Print,},%
   sensitive=true,%
   morecomment=[l]{//},%
   morecomment=[s]{/*}{*/},%
   morestring=[b]',%
   morestring=[b]",%
   morestring=[s]{`}{`},%
} 

\lstdefinelanguage{Stan}{
  morekeywords=[1]{functions,data,parameters,transformed,model,generated,quantities,%
    for,in,while,print,if,else,lower,upper,increment_log_prob,T,return,%
    reject,integrate_ode,integrate_ode_bdf,integrate_ode_rk45,target},%
  morekeywords=[2]{int,real,vector,%
    ordered,positive_ordered,simplex,unit_vector,%
    row_vector,matrix,%
    cholesky_factor_corr,cholesky_factor_cov,%
    coor_matrix,cov_matrix,%
    void},%
  morekeywords=[3]{%
    Phi,%
    Phi_approx,%
    abs,%
    acos,%
    acosh,%
    append_col,%
    append_row,%
    asin,%
    asinh,%
    atan,%
    atan2,%
    atanh,%
    bernoulli_cdf,%
    bernoulli_cdf_log,%
    bernoulli_lccdf,%
    bernoulli_lcdf,%
    bernoulli_logit_lpmf,%
    bernoulli_logit_lpmf,%
    bernoulli_lpmf,%
    bernoulli_lpmf,%
    bernoulli_rng,%
    bessel_first_kind,%
    bessel_second_kind,%
    beta_binomial_cdf,%
    beta_binomial_cdf_log,%
    beta_binomial_lccdf,%
    beta_binomial_lcdf,%
    beta_binomial_lpmf,%
    beta_binomial_lpmf,%
    beta_binomial_rng,%
    beta_cdf,%
    beta_cdf_log,%
    beta_lccdf,%
    beta_lcdf,%
    beta_lpdf,%
    beta_lpdf,%
    beta_rng,%
    binary_log_loss,%
    binomial_cdf,%
    binomial_cdf_log,%
    binomial_coefficient_log,%
    binomial_lccdf,%
    binomial_lcdf,%
    binomial_logit_lpmf,%
    binomial_logit_lpmf,%
    binomial_lpmf,%
    binomial_lpmf,%
    binomial_rng,%
    block,%
    categorical_logit_lpmf,%
    categorical_logit_lpmf,%
    categorical_lpmf,%
    categorical_lpmf,%
    categorical_rng,%
    cauchy_cdf,%
    cauchy_cdf_log,%
    cauchy_lccdf,%
    cauchy_lcdf,%
    cauchy_lpdf,%
    cauchy_lpdf,%
    cauchy_rng,%
    cbrt,%
    ceil,%
    chi_square_cdf,%
    chi_square_cdf_log,%
    chi_square_lccdf,%
    chi_square_lcdf,%
    chi_square_lpdf,%
    chi_square_lpdf,%
    chi_square_rng,%
    cholesky_decompose,%
    col,%
    cols,%
    columns_dot_product,%
    columns_dot_self,%
    cos,%
    cosh,%
    crossprod,%
    csr_extract_u,%
    csr_extract_v,%
    csr_extract_w,%
    csr_matrix_times_vector,%
    csr_to_dense_matrix,%
    cumulative_sum,%
    determinant,%
    diag_matrix,%
    diag_post_multiply,%
    diag_pre_multiply,%
    diagonal,%
    digamma,%
    dims,%
    dirichlet_lpdf,%
    dirichlet_lpdf,%
    dirichlet_rng,%
    distance,%
    dot_product,%
    dot_self,%
    double_exponential_cdf,%
    double_exponential_cdf_log,%
    double_exponential_lccdf,%
    double_exponential_lcdf,%
    double_exponential_lpdf,%
    double_exponential_lpdf,%
    double_exponential_rng,%
    e,%
    eigenvalues_sym,%
    eigenvectors_sym,%
    erf,%
    erfc,%
    exp,%
    exp2,%
    exp_mod_normal_cdf,%
    exp_mod_normal_cdf_log,%
    exp_mod_normal_lccdf,%
    exp_mod_normal_lcdf,%
    exp_mod_normal_lpdf,%
    exp_mod_normal_lpdf,%
    exp_mod_normal_rng,%
    expm1,%
    exponential_cdf,%
    exponential_cdf_log,%
    exponential_lccdf,%
    exponential_lcdf,%
    exponential_lpdf,%
    exponential_lpdf,%
    exponential_rng,%
    fabs,%
    falling_factorial,%
    fdim,%
    floor,%
    fma,%
    fmax,%
    fmin,%
    fmod,%
    frechet_cdf,%
    frechet_cdf_log,%
    frechet_lccdf,%
    frechet_lcdf,%
    frechet_lpdf,%
    frechet_lpdf,%
    frechet_rng,%
    gamma_cdf,%
    gamma_cdf_log,%
    gamma_lccdf,%
    gamma_lcdf,%
    gamma_lpdf,%
    gamma_lpdf,%
    gamma_p,%
    gamma_q,%
    gamma_rng,%
    gaussian_dlm_obs_lpdf,%
    gaussian_dlm_obs_lpdf,%
    get_lp,%
    gumbel_cdf,%
    gumbel_cdf_log,%
    gumbel_lccdf,%
    gumbel_lcdf,%
    gumbel_lpdf,%
    gumbel_lpdf,%
    gumbel_rng,%
    head,%
    hypergeometric_lpmf,%
    hypergeometric_lpmf,%
    hypergeometric_rng,%
    hypot,%
    if_else,%
    inc_beta,%
    int_step,%
    inv,%
    inv_chi_square_cdf,%
    inv_chi_square_cdf_log,%
    inv_chi_square_lccdf,%
    inv_chi_square_lcdf,%
    inv_chi_square_lpdf,%
    inv_chi_square_lpdf,%
    inv_chi_square_rng,%
    inv_cloglog,%
    inv_gamma_cdf,%
    inv_gamma_cdf_log,%
    inv_gamma_lccdf,%
    inv_gamma_lcdf,%
    inv_gamma_lpdf,%
    inv_gamma_lpdf,%
    inv_gamma_rng,%
    inv_logit,%
    inv_phi,%
    inv_sqrt,%
    inv_square,%
    inv_wishart_lpdf,%
    inv_wishart_lpdf,%
    inv_wishart_rng,%
    inverse,%
    inverse_spd,%
    is_inf,%
    is_nan,%
    lbeta,%
    lchoose,%
    lgamma,%
    lkj_corr_cholesky_lpdf,%
    lkj_corr_cholesky_lpdf,%
    lkj_corr_cholesky_rng,%
    lkj_corr_lpdf,%
    lkj_corr_lpdf,%
    lkj_corr_rng,%
    lmgamma,%
    lmultiply,%
    log,%
    log10,%
    log1m,%
    log1m_exp,%
    log1m_inv_logit,%
    log1p,%
    log1p_exp,%
    log2,%
    log_determinant,%
    log_diff_exp,%
    log_falling_factorial,%
    log_inv_logit,%
    log_mix,%
    log_rising_factorial,%
    log_softmax,%
    log_sum_exp,%
    logistic_cdf,%
    logistic_cdf_log,%
    logistic_lccdf,%
    logistic_lcdf,%
    logistic_lpdf,%
    logistic_lpdf,%
    logistic_rng,%
    logit,%
    lognormal_cdf,%
    lognormal_cdf_log,%
    lognormal_lccdf,%
    lognormal_lcdf,%
    lognormal_lpdf,%
    lognormal_lpdf,%
    lognormal_rng,%
    machine_precision,%
    max,%
    mdivide_left_tri_low,%
    mdivide_right_tri_low,%
    mean,%
    min,%
    modified_bessel_first_kind,%
    modified_bessel_second_kind,%
    multi_gp_cholesky_lpdf,%
    multi_gp_cholesky_lpdf,%
    multi_gp_lpdf,%
    multi_gp_lpdf,%
    multi_normal_cholesky_lpdf,%
    multi_normal_cholesky_lpdf,%
    multi_normal_cholesky_rng,%
    multi_normal_lpdf,%
    multi_normal_lpdf,%
    multi_normal_prec_lpdf,%
    multi_normal_prec_lpdf,%
    multi_normal_rng,%
    multi_student_t_lpdf,%
    multi_student_t_lpdf,%
    multi_student_t_rng,%
    multinomial_lpmf,%
    multinomial_lpmf,%
    multinomial_rng,%
    multiply_log,%
    multiply_lower_tri_self_transpose,%
    neg_binomial_2_cdf,%
    neg_binomial_2_cdf_log,%
    neg_binomial_2_lccdf,%
    neg_binomial_2_lcdf,%
    neg_binomial_2_log_lpmf,%
    neg_binomial_2_log_lpmf,%
    neg_binomial_2_log_rng,%
    neg_binomial_2_lpmf,%
    neg_binomial_2_lpmf,%
    neg_binomial_2_rng,%
    neg_binomial_cdf,%
    neg_binomial_cdf_log,%
    neg_binomial_lccdf,%
    neg_binomial_lcdf,%
    neg_binomial_lpmf,%
    neg_binomial_lpmf,%
    neg_binomial_rng,%
    negative_infinity,%
    normal_cdf,%
    normal_cdf_log,%
    normal_lccdf,%
    normal_lcdf,%
    normal_lpdf,%
    normal_lpdf,%
    normal_rng,%
    not_a_number,%
    num_elements,%
    ordered_logistic_lpmf,%
    ordered_logistic_lpmf,%
    ordered_logistic_rng,%
    owens_t,%
    pareto_cdf,%
    pareto_cdf_log,%
    pareto_lccdf,%
    pareto_lcdf,%
    pareto_lpdf,%
    pareto_lpdf,%
    pareto_rng,%
    pareto_type_2_cdf,%
    pareto_type_2_cdf_log,%
    pareto_type_2_lccdf,%
    pareto_type_2_lcdf,%
    pareto_type_2_lpdf,%
    pareto_type_2_lpdf,%
    pareto_type_2_rng,%
    pi,%
    poisson_cdf,%
    poisson_cdf_log,%
    poisson_lccdf,%
    poisson_lcdf,%
    poisson_log_lpmf,%
    poisson_log_lpmf,%
    poisson_log_rng,%
    poisson_lpmf,%
    poisson_lpmf,%
    poisson_rng,%
    positive_infinity,%
    pow,%
    prod,%
    qr_Q,%
    qr_R,%
    quad_form,%
    quad_form_diag,%
    quad_form_sym,%
    rank,%
    rayleigh_cdf,%
    rayleigh_cdf_log,%
    rayleigh_lccdf,%
    rayleigh_lcdf,%
    rayleigh_lpdf,%
    rayleigh_lpdf,%
    rayleigh_rng,%
    rep_array,%
    rep_matrix,%
    rep_row_vector,%
    rep_vector,%
    rising_factorial,%
    round,%
    row,%
    rows,%
    rows_dot_product,%
    rows_dot_self,%
    scaled_inv_chi_square_cdf,%
    scaled_inv_chi_square_cdf_log,%
    scaled_inv_chi_square_lccdf,%
    scaled_inv_chi_square_lcdf,%
    scaled_inv_chi_square_lpdf,%
    scaled_inv_chi_square_lpdf,%
    scaled_inv_chi_square_rng,%
    sd,%
    segment,%
    sin,%
    singular_values,%
    sinh,%
    size,%
    skew_normal_cdf,%
    skew_normal_cdf_log,%
    skew_normal_lccdf,%
    skew_normal_lcdf,%
    skew_normal_lpdf,%
    skew_normal_lpdf,%
    skew_normal_rng,%
    softmax,%
    sort_asc,%
    sort_desc,%
    sort_indices_asc,%
    sort_indices_desc,%
    sqrt,%
    sqrt2,%
    square,%
    squared_distance,%
    step,%
    student_t_cdf,%
    student_t_cdf_log,%
    student_t_lccdf,%
    student_t_lcdf,%
    student_t_lpdf,%
    student_t_lpdf,%
    student_t_rng,%
    sub_col,%
    sub_row,%
    sum,%
    tail,%
    tan,%
    tanh,%
    tcrossprod,%
    tgamma,%
    to_array_1d,%
    to_array_2d,%
    to_matrix,%
    to_row_vector,%
    to_vector,%
    trace,%
    trace_gen_quad_form,%
    trace_quad_form,%
    trigamma,%
    trunc,%
    uniform_cdf,%
    uniform_cdf_log,%
    uniform_lccdf,%
    uniform_lcdf,%
    uniform_lpdf,%
    uniform_lpdf,%
    uniform_rng,%
    variance,%
    von_mises_lpdf,%
    von_mises_lpdf,%
    von_mises_rng,%
    weibull_cdf,%
    weibull_cdf_log,%
    weibull_lccdf,%
    weibull_lcdf,%
    weibull_lpdf,%
    weibull_lpdf,%
    weibull_rng,%
    wiener_lpdf,%
    wiener_lpdf,%
    wishart_lpdf,%
    wishart_lpdf,%
    wishart_rng
  },%
  otherkeywords={<-,~,+=,=},%
  sensitive=true,%
  morecomment=[l]{\#},%
  morecomment=[l]{//},%
  morecomment=[n]{/*}{*/},%
  string=[d]"
  literate={<-}{{$\leftarrow$}}1 {~}{{$\sim$}}1%
}

\acmConference[Onward! 2019]{SPLASH Onward! 2019}{October 20-25, 2019}{Athens, Greece}

\setcopyright{none}

\title{Deployable Probabilistic Programming}
\author{David Tolpin}
\affiliation{
    \institution{PUB+}
    \country{Israel}
}
\email{david.tolpin@gmail.com}

\begin{abstract}
  We propose design guidelines for a probabilistic programming
  facility suitable for deployment as a part of a production
  software system. As a reference implementation, we introduce
  Infergo, a probabilistic programming facility for Go, a modern
  programming language of choice for server-side software
  development. We argue that a similar probabilistic
  programming facility can be added to most modern general-purpose
  programming languages.

  Probabilistic programming enables automatic tuning of
  program parameters and algorithmic decision making through
  probabilistic inference based on the data. To facilitate
  addition of probabilistic programming capabilities to other
  programming languages, we share implementation choices and
  techniques employed in development of Infergo. We illustrate
  applicability of Infergo to various use cases on case
  studies, and evaluate Infergo's performance on several
  benchmarks, comparing Infergo to dedicated inference-centric
  probabilistic programming frameworks.
\end{abstract}

\begin{document}
\maketitle

\begin{sloppypar}

\section{Introduction}

  Probabilistic programming \cite{GMR+08,MSP14,WVM14,GS15}
  represents statistical models as programs written in an
  otherwise general programming language that provides syntax
  for the definition and conditioning of random variables.
  Inference can be performed on probabilistic programs to
  obtain the posterior distribution or point estimates of the
  variables. Inference algorithms are provided by the
  probabilistic programming framework, and each algorithm is
  usually applicable to a wide class of probabilistic programs
  in a black-box manner. The algorithms include
  Metropolis-Hastings\cite{WSG11,MSP14,YHG14}, Hamiltonian
  Monte Carlo~\cite{Stan17}, expectation
  propagation~\cite{MWG+10}, extensions of Sequential Monte
  Carlo~\cite{WVM14,MYM+15,PWD+14,RNL+2016}, variational
  inference~\cite{WW13,KTR+17}, gradient-based
  optimization~\cite{Stan17,BCJ+19}, and others.

  There are two polar views on the role of probabilistic
  programming.  One view is that probabilistic programming is
  a flexible framework for specification and analysis of
  statistical models~\cite{WSG11,TMY+16,GXG18}. Proponents of
  this view perceive probabilistic programming as a tool for
  data science practitioners. A typical workflow consists of
  data acquisition and pre-processing, followed by several
  iterations of exploratory model design and testing of
  inference algorithms. Once a sufficiently robust statistical
  model is obtained, analysis results are post-processed,
  visualized, and occasionally integrated into algorithms of a
  production software system. 

  The other, emerging, view is that probabilistic programming
  is an extension of a regular programming toolbox allowing
  algorithms implemented in general-purpose programming
  languages to have learnable parameters. In this view,
  statistical models are integrated into the software system,
  and inference and optimization take place during data
  processing, prediction, and algorithmic decision making in
  production.  Probabilistic programming code runs unattended,
  and inference results are an integral part of the
  algorithms. Natural applications arise whenever the algorithm
  is a generative model of a mental or physical process, in
  particular when latent variables of the algorithm are used for
  decision making~\cite{GT+16,WBD+19}; for example, inference
  about user behavior in social networks, ongoing health
  monitoring, or online motion planning of autonomous robotic
  devices.

  In this work we are concerned with the later view on
  probabilistic programming. Our objective is to pave a path
  to deployment of probabilistic programs in production
  systems, promoting a wider adoption of probabilistic
  programming as a flexible tool for ubiquitous statistical
  inference. Most probabilistic programming frameworks are
  suited, to a certain extent, to be used in either the
  exploratory or the production scenario. However, the
  proliferation of probabilistic programming languages and
  frameworks \footnote{There are 45 probabilistic programming
  languages listed on
  Wikipedia~\cite{wiki:Probabilistic_programming_language}. 18
  probabilistic programming languages were presented during
  the developer meetup at PROBPROG'2018, the inaugural
  conference on probabilistic
  programming,~\url{http://probprog.cc/}, half of which are
  not on the Wikipedia list.} on one hand, and scarcity of success
  stories about production use of probabilistic programming on
  the other hand, suggest that there is a need for new ideas
  and approaches to make probabilistic programming models
  available for production use.

  Our attitude differs from the established approach in that
  instead of proposing yet another probabilistic programming
  language, either genuinely different~\cite{MMR+07,Stan17} or
  derived from an existing general-purpose programming
  language by extending the syntax or changing the
  semantics~\cite{GMR+08,TMY+16,GXG18}, we advocate enabling
  probabilistic inference on models written in a
  general-purpose probabilistic programming language, the same
  one as the language used to program the bulk of the system. 
  We formulate guidelines for such implementation, and, based
  on the guidelines, introduce a probabilistic programming
  facility for the Go programming language~\cite{Golang}. By
  explaining our implementation choices and techniques,
  we argue that a similar facility can be added to any modern
  general-purpose purpose programming language, giving a
  production system modeling and inference capabilities which
  do not fall short from and sometimes exceed those of
  probabilistic programming-centric frameworks with custom
  languages and runtimes. Availability of such facilities
  in major general-purpose programming languages would free
  probabilistic programming researchers and practitioners
  alike from language wars~\cite{SH14} and foster practical
  applications of probabilistic programming.

  \subsection*{Contributions}

  This work brings the following major contributions:

  \begin{itemize}
    \item Guidelines for design and implementation of a
      probabilistic programming facility deployable as a part 
      of a production software system.
    \item A reference implementation of deployable probabilistic
      programming facility for the Go programming language.
    \item Implementation highlights outlining important
      choices we made in implementation of the
      probabilistic programming facility, and providing a
      basis for implementation of a similar facility for
      other general-purpose programming languages.
  \end{itemize}

\section{Related Work}

Our work is related to three interconnected areas of research:
\begin{enumerate}
  \item Design and implementation of probabilistic programming
    languages. 
  \item Integration and interoperability between probabilistic
    programming models and the surrounding software
    systems.
  \item Differentiable programming for machine learning.
\end{enumerate}

Probabilistic programming is traditionally associated with
design and implementation of probabilistic programming
\textit{languages}. An apparent justification for a new
programming language is that a probabilistic program has a
different semantics~\cite{GHNR14,SYH+16} than a `regular'
program of similar structure. \citet{GS15} offer a systematic
approach to design and implementation of probabilistic
programming languages based on extension of the syntax of a
(subset of) general-purpose programming language and
transformational compilation to continuation-passing
style~\cite{AJ89,A07}.  \citet{MPY+18} give a comprehensive overview
of available choices of design and implementation of first-order
and higher-order probabilistic programming languages.
Stan~\cite{Stan17} is built around an imperative probabilistic
programming language with program structure tuned for most
efficient execution of inference on the model.
SlicStan~\cite{GGS19} is built on top of Stan as a
source-to-source compiler, providing the model developer with a
more intuitive language while retaining performance benefits of
Stan. Our work differs from existing approaches in that we
advocate enabling probabilistic programming in a general-purpose
programming language by leveraging existing capabilities of the
language, rather than building a new language on top or besides
an existing one.

The need for integration between probabilistic programs and the
surrounding software environment is well understood in the
literature~\cite{TMY+16,BCJ+19}. Probabilistic programming
languages are often implemented as embedded domain-specific
languages~\cite{SGG15,GXG18,TMY+16} and include a mechanism of
calling functions from libraries of the host languages. Our work
goes a step further in this direction: since the language of
implementation of probabilistic models coincides with the host
language, any library or user-defined function of the host
language can be directly called from the probabilistic model,
and model methods can be directly called from the host language.

Automatic differentiation is widely employed in machine
learning~\cite{BPR+17,MBB+18} where it is also known as
`differentiable programming', and is responsible for enabling
efficient inference in many probabilistic programming
frameworks~\cite{Stan17,GXG18,ISF+18,BCJ+19,THS+17}. Different
automatic differentiation techniques~\cite{GW08} allow 
different compromises between flexibility, efficiency, and
feature-richness~\cite{SDT14,PGC+17,ISF+18}.  Automatic
differentiation is usually implemented through either
operator overloading~\cite{Stan17,PGC+17,GXG18} or source
code transformation~\cite{ISF+18,MMW18}. Our work, too, relies
on automatic differentiation, implemented as source code
transformation. However, a novelty of our approach is that
instead of using explicit calls or directives to denote parts of
code which need to be differentiated, we rely on the type
system of Go to selectively differentiate the code relevant for
inference, thus combining advantages of both operator
overloading and source code transformation.

\section{Challenges}
\label{sec:challenges}

Incorporating a probabilistic program, or rather a probabilistic
procedure, within a larger code body appears to be rather
straightforward: one implements the model in the probabilistic
programming language, fetches and preprocesses the data in the
host programming language, passes the data and the model to an
inference algorithm, and post-processes the results in the host
programming language again to make algorithmic decisions based
on inference outcomes. To choose the best option for each
particular application, probabilistic programming languages and
frameworks are customarily compared by the expressive power
(classes of probabilistic models that can be represented),
conciseness (e.g.  the number of lines of code to describe a
particular model), and time and space efficiency of inference
algorithms~\cite{WVM14,P16,R17,GXG18}.  However, complex
software systems make integration of probabilistic inference
challenging, and considerations beyond expressiveness
and performance come into play.

\paragraph{Simulation vs. inference} Probabilistic models often
follow the design pattern of simulation-inference: a significant
part of the model is a simulator running an algorithm with
given parameters; the optimal parameters, or their distribution,
are to be inferred. The inferred parameters are then used by the
software system to execute the simulation independently of
inference for prediction and decision making.

This pattern suggests re-use of the simulator: instead of
implementing the simulator twice, in the probabilistic model and
in the host environment, one can invoke the same code for both
purposes.  However to achieve this, the host language must
coincide with the implementation language of the probabilistic
model, on one hand, and allow a computationally efficient
implementation of the simulation, on the other hand. Some
probabilistic programming frameworks (Figaro~\cite{P09},
Anglican~\cite{TMY+16}, Turing~\cite{GXG18}) are built with
tight integration with the host environment in mind; more often
than not though the probabilistic code is not trivial to re-use.

\paragraph{Data interface} The data for inference may come from
a variety of sources: network, databases, distributed file
systems, and in many different formats. Efficient inference
depends on fast data access and updating. Libraries for data
access and manipulation are available in the host environment.
While the host environment can be used as a proxy retrieving and
transforming the data, such as in the case of Stan~\cite{Stan17}
integrations, sometimes direct access from the probabilistic
code is the preferred option, for example when the data is
streamed or retrieved conditionally, as in active
learning~\cite{SLB15}.

A flexible data interface is also beneficial for support of
mini-batch optimization~\cite{LZC+14}. Mini-batch optimization
is used when the data set is too big to fit in memory, and the
objective function gradient is estimated based on mini-batches
--- small portions of data. Different models and inference
objectives may require different mini-batch loading schemes.
For best performance of mini-batch optimization, it should be
possible to program loading of mini-batches on case-by-case
basis.

\paragraph{Integration and deployment} Deployment of software
systems is a delicate process involving automatic builds and
maintenance of dependencies. Adding a component, which possibly
introduces additional software dependencies or even a separate
runtime, complicates deployment. Minimizing the burden of
probabilistic programming on the integration and deployment
process should be a major consideration in design or selection
of probabilistic programming tools. Probabilistic programming
frameworks that are implemented or provide an interface in a
popular programming language, e.g.  Python
(Edward~\cite{THS+17}, Pyro~\cite{BCJ+19}) are easier to
integrate and deploy, however the smaller the footprint of a
probabilistic programming framework, the easier is the adoption.
An obstacle in integration of probabilistic programming lies
also in adoption of the probabilistic programming
language~\cite{MR12}, if the latter differs from the
implementation language of the rest of the system.

\section{Guidelines}

Based on the experience of developing and deploying solutions
using different probabilistic programming environments, we came
up with guidelines to implementation of a probabilistic
programming facility for server-side applications. We believe
that these guidelines, when followed, help easier integration of
probabilistic programming inference into large-scale software
systems.

\begin{enumerate}
\item A probabilistic model should be programmed in the host
programming language. The facility may impose a discipline on
model implementation, such as through interface constraints, but
otherwise supporting unrestricted use of the host language for
implementation of the model.

\item Built-in and user-defined data structures and libraries
should be accessible in the probabilistic programming model.
Inference techniques relying on the code structure, such as
those based on automatic differentiation, should support the
use of common data structures of the host language.

\item The model code should be reusable between inference and
simulation. The code which is not required solely for inference
should be written once for both inference of parameters and use
of the parameters in the host environment.  It should be
possible to run simulation outside the probabilistic model without
runtime or memory overhead imposed by inference needs.
\end{enumerate}

\section{Probabilistic Programming in Go}

In line with the guidelines, we have implemented a probabilistic
programming facility for the Go programming language, Infergo.
We have chosen Go because Go is a
small but expressive programming language with efficient
implementation that has recently become quite popular for
computation-intensive server-side programming. Infergo is used
in production environment for inference of mission-critical
algorithm parameters.

\subsection{An Overview of Infergo}

Infergo comprises a command-line tool that augments the source
code of a model to facilitate inference, a collection of basic
models (distributions) serving as building blocks for other
models, and a library of inference algorithms.

A probabilistic model in Infergo is an implementation of an
interface requiring a single method
\lstinline{Observe}\footnote{The method name follows
Venture~\cite{MSP14}, Anglican~\cite{TMY+16}, and other
probabilistic programming languages in which \textit{observe} is
the language construct for conditioning of the model on
observations.} accepts a vector (a Go \textit{slice}) of floats,
the parameters to infer, and returns a single float, interpreted
as unnormalized log-likelihood of the posterior distribution.
The model object usually encapsulates the observations, either
as a data field or as a reference to a data source.
Implementation of model methods can be written in virtually
unrestricted Go and use any Go libraries.

For inference, Infergo relies on automatic differentiation. The
source code of the model is translated by the command-line tool
into an equivalent model with reverse-mode automatic
differentiation of the log-likelihood with respect to the
parameters applied. The differentiation operates on the built-in
floating-point type and incurs only a small computational
overhead. However, even this overhead is avoided when the model
code is executed outside inference algorithms: both the original
and the differentiated model are simultaneously available to the
rest of the program code, so the methods can be called on the
differentiated model for inference, and on the original model
for the most efficient execution with the inferred parameters.

\subsection{A Basic Example}
\label{sec:basic-example}

Let us illustrate the use of Infergo on a basic example serving
as a probabilistic programming equivalent of the "Hello, World!"
program --- inferring parameters of a unidimensional Normal
distribution. In this example, the model object holds the set of
observations from which the distribution parameters are to be
inferred:
\begin{lstlisting}
type ExampleModel struct {
  Data []float64
}
\end{lstlisting}

The \lstinline{Observe} method computes the log-likelihood of
the distribution parameters. The Normal distribution has two
parameters: mean $\mu$ and standard deviation $\sigma$. To ease
the inference, positive $\sigma$ is transformed to unrestricted
$\log \sigma$, and the parameter vector \lstinline{x} is $[\mu,
\log \sigma]$. \lstinline{Observe} first imposes a prior on the
parameters, and then conditions the posterior distribution on
the observations:

\begin{lstlisting}
// x[0] is the mean,
// x[1] is the log stddev of the distribution
func (m *ExampleModel)
     Observe(x []float64) float64 {
  // Our prior is a unit normal ...
  ll := Normal.Logps(0, 1, x...)
  // ... but the posterior is based 
  // on data observations.
  ll += Normal.Logps(x[0], math.Exp(x[1]),
                     m.Data...)
  return ll
}
\end{lstlisting}

For inference, we supply the data and optionally initialize the
parameter vector. Here the data is hard-coded, but in general
can be read from a file, a database, a network connection,
or dynamically generated by another part of the program:

\begin{lstlisting}
// Data
m := &ExampleModel{[]float64{
  -0.854, 1.067, -1.220, 0.818, -0.749,
  0.805, 1.443, 1.069, 1.426, 0.308}}

// Parameters
x := []float64{}
\end{lstlisting}
  
The fastest and most straightforward inference is the maximum
\textit{a posteriori} (MAP) estimation, which can be done using
a gradient descent algorithm (Adam~\cite{KB15} in this case):

\begin{lstlisting}
opt := &infer.Adam{
  Rate:  0.01,
}
for iter := 0; iter != 1000; iter++ {
  opt.Step(m, x)
}
\end{lstlisting}

To recover the full posterior distribution of
the parameters we use Hamiltonian Monte Carlo~\cite{N12}:

\begin{lstlisting}
hmc := &infer.HMC{
  Eps: 0.1,
}
samples := make(chan []float64)
hmc.Sample(m, x, samples)
for i := 0; i != 5000; i++ {
  x = <-samples
}
hmc.Stop()
\end{lstlisting}

Models are of course not restricted to linear code, and may
comprise multiple methods containing loops, conditional
statements, function and method calls, and recursion. Case
studies~(Section~\ref{sec:case-studies}) provide more
model examples.

\subsection{Model Syntax and Rules of Differentiation}

In Infergo, the inference is performed on a model. Just like in
Stan~\cite{Stan17} models, latent random variables in Infergo
models are continuous. Additionally, Infergo can express and
perform inference on non-deterministic models by including
stochastic choices in the model code.

An Infergo model is a data type and methods defined on the data
type.  Inference algorithms accept a model object as an
argument.  A model object usually encapsulates observations ---
the data to which the parameter values are adapted.

A model is implemented in Go, virtually
unrestricted\footnote{There are insignificant implementation
limitations which are gradually lifted as Infergo is being
developed.}. Because of this, the rules of model specification
and differentiation are quite straightforward. Nonetheless, the
differentiation rules ensure that provided that the model's
\lstinline{Observe} method (including calls to other model methods)
is differentiable, the gradient is properly computed~\cite{GW08}.

A model is defined in its own package. The model must
implement interface \lstinline{Model} containing a single method
\lstinline{Observe}. \lstinline{Observe} accepts a slice of
\lstinline{float64} --- the model parameters --- and returns a
\lstinline{float64} scalar. For probabilistic inference, the
returned value is interpreted as the log-likelihood of the model
parameters. Inference models rely on computing the gradient of
the returned value with respect to the model parameters through
automatic differentiation. In the model's source code:

\begin{enumerate}
  \item Methods on the type implementing \lstinline{Model} returning a
    single \lstinline{float64} or nothing are differentiated.
  \item Within the methods, the following is differentiated:
    \begin{itemize}
      \item assignments to \lstinline{float64} (including parallel
        assignments if all values are of type
        \lstinline{float64});
      \item returns of \lstinline{float64};
      \item standalone calls to methods on the type implementing
        \lstinline{Model} (apparently called for side  effects on
        the model).
    \end{itemize}
\end{enumerate}

Functions for which the gradient is provided rather than derived
through automatic differentiation are called
\textit{elementals}~\cite{GW08}.  Derivatives do not propagate
through a function that is not an elemental or a call to a model
method. If a derivative is not registered for an elemental,
calling the elemental in a differentiated context will cause a
run-time error.

Functions are considered elementals (and must have a
registered derivative) if their signature is of kind
\lstinline{func (float64, float64*) float64}; that is, 
one or more non-variadic \lstinline{float64} arguments and
\lstinline{float64} return value. For example, function
\lstinline{func (float64, float64, float64) float64}
is considered elemental, while functions
\begin{itemize}
  \item \lstinline{func (...float64) float64}
  \item \lstinline{func ([]float64) float64}
  \item \lstinline{func (int, float64) float64}
\end{itemize}
are not. Gradients for selected functions from the
\lstinline{math} package are pre-defined (\lstinline{Sqrt},
\lstinline{Exp}, \lstinline{Log}, \lstinline{Pow},
\lstinline{Sin}, \lstinline{Cos}, \lstinline{Tan}). Auxiliary
elemental functions with pre-defined gradients are provided in a
separate package.

Any user-defined function of appropriate kind can be used as an
elemental inside model methods. The gradient must be registered
for the function using a call to
\begin{lstlisting}
func RegisterElemental(
  f interface{}, 
  g func(value float64,
         params ...float64) []float64)
\end{lstlisting} For
example, the call
\begin{lstlisting}
RegisterElemental(Sigm,
  func(value float64,
       _ ...float64) []float64 {
    return []float64{value*(1-value)}
  })
\end{lstlisting}
registers the gradient for function
\lstinline{Sigm}$(x)\coloneq\left(1+e^{-x}\right)^{-1}$.

\subsection{Inference}

Out of the box, Infergo offers 
\begin{itemize}
  \item maximum \textit{a posteriori} (MAP) estimation by stochastic
    gradient descent~\cite{R16}; 
  \item full posterior inference by Hamiltonian Monte
    Carlo~\cite{N12,HG11}.
\end{itemize}

When only a point estimate of the model parameters is required,
as, for example, in the simulation-inference case, stochastic
gradient descent is a fast and robust inference family. Infergo
offers stochastic gradient descent with momentum and
Adam~\cite{KB15}.  The inference is performed by calling the
\lstinline{Step} method of the optimizer until a termination
condition is met:
\begin{lstlisting}
for !<termination condition> {
    optimizer.Step(model, parameters)
}
\end{lstlisting}

When the full posterior must be recovered for integration,
Hamiltonian Monte Carlo (HMC) can be used instead. Vanilla HMC
as well as the No-U-Turn sampler (NUTS)~\cite{HG11} are
provided. To support lazy any-time computation, the inference
runs in a separate \textit{goroutine}, connected by a channel to
the caller, and asynchronously writes samples to the channel.
The caller reads as many samples from the channel as needed, and
further postprocesses them:
\begin{lstlisting}
samples := make(chan []float64)
hmc.Sample(model, parameters, samples)
for !<termination condition> {
  sample = <-samples
  // postprocess the sample
}
hmc.Stop()
\end{lstlisting}

Other inference schemes, most notably automatic differentiation
variational inference~\cite{KTR+17}, are planned for addition
in future versions of Infergo. However, an inference algorithm
does not have to be a part of Infergo to work with Infergo
models. Since Infergo models operate on a built-in Go floating
point type, \lstinline{float64}, any third-party inference or
optimization library can be used as well. 

As an example, the Gonum~\cite{Gonum} library for numerical
computations contains an optimization package offering several
algorithms for gradient-based optimization. A Gonum optimization
algorithm requires the function to optimize, as well as the
function's gradient, as parameters. An Infergo model can be
trivially wrapped for use with Gonum, and Infergo provides a
convenience function
\begin{lstlisting}
func FuncGrad(m Model) (
  Func func(x []float64) float64,
  Grad func(grad []float64, x []float64))
\end{lstlisting}
which accepts a model and returns the model's
\lstinline{Observe} method and its gradient suitable for passing
to Gonum optimization algorithms. Among available algorithms are
advanced versions of stochastic gradient descent, as well as
algorithms of BFGS family, in particular L-BFGS~\cite{LN89},
providing for a more efficient MAP estimation than stochastic
gradient descent at the cost of higher memory consumption:
\begin{lstlisting}
function, gradient := FuncGrad(model)
problem := gonum.Problem{Func: function, 
                         Grad: gradient}
result, err := gonum.Minimize(problem, parameters)
\end{lstlisting}

Other third-party implementations of optimization algorithms can
be used just as easily, thanks to Infergo models being pure Go
code operating on built-in Go data types.

\section{Implementation Highlights}

In our journey towards implementation and deployment of Infergo, we
considered options and made implementation decisions which we
believe to have been crucial for addressing the challenges and
following the guidelines we formulated for implementation of a
deployable probabilistic programming facility. While different
choices may be as well suited for the purpose as the ones we
made, we feel it is important to share our contemplations and
justify decisions, to which the rest of this section is
dedicated, from choosing the implementation language, through
details of automatic differentiation, to support for inference
on streaming and stochastic observations.

\subsection{Choice of Programming Language}

Python~\cite{Python95}, Scala~\cite{Scala06}, and
Julia~\cite{Julia14} have been popular choices for development
of probabilistic programming languages and frameworks integrated
with general-purpose programming
languages~\cite{S16,MMW18,BCJ+19,P09,Rainier,GXG18,ISF+18}.
Each of these languages has advantages for implementing a
probabilistic programming facility. However, for a reference
implementation of a facility best suited for probabilistic
inference in a production software system, we were looking
for a language that \begin{itemize}
  \item compiles into fast and compact code;
  \item does not incur dependency on a heavy runtime; and
  \item is a popular choice for building server-side software.
\end{itemize}
In addition, since we were going to automatically differentiate
the language's source code, we were looking for a relatively
small language, preferably one for which parsing and program
analysis tools are readily available. 

We chose Go~\cite{Golang}. Go is a small but expressive
programming language widely used for implementing
computationally-intensive server-side software systems. 
The standard libraries and development environment offer
capabilities which made implementation of Infergo affordable
and streamlined:

\begin{itemize}
  \item The Go parser and abstract syntax tree serializer are
    a part of the standard library. Parsing, transforming,
    and generating Go source code is straightforward and
    effortless.
  \item Type inference (or \textit{type checking} as it is
    called in the Go ecosystem), also provided in the
	standard library, augments parsing, and allows to
    selectively apply transformation-based automatic
    differentiation  based on static expression types. 
  \item Go provides reflection capabilities for almost any aspect
    of the running program, without compromising efficiency.
    Reflection gives the flexibility highly desirable to
    implement features such as calling Go library
    functions from the model code.
  \item Go compiles and runs fast. Fast compilation and
    execution speeds allow to use the same facility for both
    exploratory design of probabilistic models and for
    inference in production environment.
  \item Go is available on many platforms and for a wide range
    of operating systems and environments, from portable
    devices to supercomputers. Go programs can be
	cross-compiled and reliably developed and deployed in
	heterogeneous environments.
  \item Go offers efficient parallel execution as a
    first-class feature, via so-called \textit{goroutines}.
    Goroutines streamline implementation of sampling-based
    inference algorithms. Sample generators and consumers
    are run in parallel, communicating through channels. 
    Inference is easy to parallelize in order to exploit
    hardware multi-processing, and samples are retrieved
    lazily for postprocessing. 
\end{itemize}

Considering and choosing Go for development of what later became
Infergo also affected the shape of the software system in which
Infergo was deployed first. Formerly implemented mostly in
Python, and both enjoying a wide choice of libraries for machine
learning and data processing and suffering from limitations of
Python as a language for implementation of computation-intensive
algorithms, the system has been gradually drifting towards
adopting Go, with performance-critical parts re-implemented and
running faster due to better hardware and network utilization,
and simpler and more maintainable code for data processing,
analysis, and decision making.

\subsection{Automatic Differentiation}

Infergo employs reverse-mode automatic differentiation via
source code transformation~\cite{GW08}. The reverse mode is the
default choice in machine learning applications, because it
works best with differentiating a scalar return value with
respect to multiple parameters. Automatic differentiation is
usually implemented using either operator overloading or source
code transformation. Operator overloading is often the method of
choice~\cite{Stan17,GXG18,PGC+17} due to relative simplicity,
but the differentiated code must use a special type (on which
the operators are overloaded) for representing floating point
numbers. Regardless of the issue of computational overhead, this
restricts the language used to specify the model to operations
on that special type and impairs interoperability: library
functions cannot be directly called but have to use an adaptor,
or a `lifted' version of the standard library must be supplied.
Similarly, the data must be cast into the special type before it
is passed to the model for inference.

\paragraph{Reverse mode source code transformation} Source code
transformation allows to automatically differentiate program
code operating on the built-in floating point type. This method
involves generation of new code, for the function gradient or
for the function itself, ahead of time or on the fly. To achieve
the best performance, some implementations generate static code
for computing the function gradient~\cite{MMW18,I18}. This
requires a rather elaborated source code analysis, and although
a significant progress has been made in this direction, not all
language constructs are supported. Alternatively, the code of
the function itself can be transformed by augmenting arithmetics
and function calls on the floating point type by function calls
that record the computations on the \textit{tape}, just like in
the case of operator overloading. Then, the backward pass is
identical to that of operator overloading. This method allows
the use of the built-in floating point type along with
virtually unrestricted language for model specification, at the
cost of slight performance loss.

This latter method is used in Infergo: the model is written in Go
and is placed in a separate Go package. A command-line tool
is run on the package to generate a sub-package containing a
differentiated version of the model.  Both the original and the
differentiated model are available simultaneously in the calling
code, so that model methods that do not require computation of
gradients can be invoked without the overhead of automatic
differentiation.  However, the performance loss is rarely a
major issue --- differentiated versions of model methods are
only 2--4 times slower on models from the case studies
(Section~\ref{sec:case-studies}), and the backward pass, which
computes the gradient, is roughly as fast as the original,
undifferentiated version. At least partially, fast backward pass
is due to careful implementation and thorough profiling of the
tape.

\paragraph{Automatic differentiation of models}
It may seem desirable to implement or use an automatic
differentiation library that is more general than needed for
probabilistic programming. However the use of such library, in
particular with source code transformation, may make
implementation restricted and less efficient: first, if a model
comprises several functions, all functions must have a signature
suitable for differentiation (that is, accept and return
floating point scalars or vectors); second, the gradient of
every nested call must be fully computed first and then used to
compute the gradient of the caller. In Infergo, only the
gradient of \lstinline{Observe} is needed, and any other
differentiated model method can only be called from
\lstinline{Observe} or another differentiated method.
Consequently, only the gradient of \lstinline{Observe} is
computed explicitly, and other differentiated methods may have
arbitrary arguments. In particular, derivatives may propagate
through fields of the model object and structured arguments.

\subsection{Model Composition} 

Just like it is possible to compose library and user-defined
functions into more complex functions, it should be possible to
compose models to build other models~\cite{THS+17,SSW+18}.
Infergo supports model composition organically: a differentiated
model method can be called from another differentiated method,
not necessarily a method of the same model object or type.
Derivatives can propagate through methods of different models
without restriction. Each model is a building block for another
model. For example, if there are two independent
single-parameter models \lstinline{A} and \lstinline{B}
describing the data, then the product model \lstinline{AB} can
be obtained by composition of \lstinline{A} and \lstinline{B}
(Figure~\ref{fig:composition}).
\begin{figure}
  \begin{lstlisting}[framexleftmargin=10pt,numbers=left]
type A struct {Data []float64}
func (model A) Observe(x []float64) {
  ...
}

type B struct {Data []float64}
func (model B) Observe(x []float64) {
  ...
}

type AB struct {Data []float64}
func (model AB) Observe(x []float64) float64 {
  return A{model.Data}.Observe(x[:1]) + 
    B{model.Data}.Observe(x[1:])
}
  \end{lstlisting}
  \caption{Model composition. Model \lstinline{AB} is the
  product of \lstinline{A} and \lstinline{B}.}
  \label{fig:composition}
\end{figure}
Combinators such as product, mixture, or hierarchy can be
realized as functions operating on Infergo models.

\paragraph{Distributions} A probabilistic programming facility
should provide a library of distributions for definition and
conditioning of random variables. To use a distribution in a
differentiable model, the distribution's $\log$ probability
density function must be differentiable by both the
distribution's parameters and by the value of the random
variable. Distributions are usually supplied as a part of the
probabilistic programming framework~\cite{TMY+16,Stan17,GXG18},
and, if at all possible, adding a user-defined distribution
requires programming at a lower level of abstraction than while
specifying a probabilistic model~\cite{S16,TMY+16,AMP17}.

In Infergo, distributions are just models, and adding and using
a custom distribution is indistinguishable from adding a model
and calling the model's method from another model. By
convention, library distributions, in addition to
\lstinline{Observe}, also have methods \lstinline{Logp} and
\lstinline{Logps} for the scalar and vector version of $\log$
probability density functions, for convenience of use in models.
The distribution library in Infergo is just a model package
which is automatically differentiated like any other model.  New
distributions can be added in any package.
Figure~\ref{fig:exponential-distribution} shows the Infergo
definition of the exponential distribution.
\begin{figure}
  \begin{lstlisting}[framexleftmargin=10pt,numbers=left]
// Exponential distribution.
type expon struct{}

// Exponential distribution, singleton instance.
var Expon expon

// Observe implements the Model interface. The
// parameter vector is lambda, observations.
func (dist expon) Observe(x []float64)
  float64 {
  lambda, y := x[0], x[1:]
  if len(y) == 1 {
    return dist.Logp(lambda, y[0])
  } else {
    return dist.Logps(lambda, y...)
  }
}

// Logp computes the log pdf of a single
// observation.
func (_ expon) Logp(lambda float64, y float64)
  float64 {
  logl := math.Log(lambda)
  return logl - lambda*y
}

// Logps computes the log pdf of a vector
// of observations.
func (_ expon) Logps(lambda float64, y ...float64)
  float64 {
  ll := 0.
  logl := math.Log(lambda)
  for i := range y {
    ll += logl - lambda*y[i]
  }
  return ll
}
  \end{lstlisting}
  \caption{The exponential distribution in Infergo.}
  \label{fig:exponential-distribution}
\end{figure}

\subsection{Stochasticity and Streaming}

Perhaps somewhat surprisingly, most probabilistic programs are
semantically deterministic~\cite{SYW+16} --- probabilistic programs
define distributions, and although they manipulate random
variables, there is no randomization or non-determinism
\textit{per se}. Randomization arises in certain inference
algorithms (notably of the Monte Carlo family) for tractable
approximate inference, but this does not introduce
non-determinism into the programs themselves. Non-deterministic
probabilistic programs arise in applications. For example,
\citet{MPT+16} explored using probabilistic programming for
policy search in non-deterministic domains. To represent
non-determinism, appropriate constructs had to be added to the
probabilistic programming language used in the study.

In Infergo, non-determinism can be easily introduced through
data. In the basic example if Section~\ref{sec:basic-example},
the data is encapsulated in the model as a vector, populated
before inference. However, this is not at all required by
Infergo.  Instead of being stored in a Go slice, the data can be
read from a Go channel or network connection and generated by
another goroutine or process.
Figure~\ref{fig:basic-example-channel} shows a modified model
from the basic example with the data gradually read from a
channel rather than passed as a whole in a field of the model
object.
\begin{figure}
  \begin{lstlisting}[framexleftmargin=10pt,numbers=left]
type StreamingModel struct {
  Data chan float64  // data is a channel
  N    int           // batch size
}

func (m *StreamingModel)
     Observe(x []float64) float64 {
  ll := Normal.Logps(0, 1, x...)
  // observe a batch of data from the channel
  for i := 0; i != m.N; i++ {
    ll += Normal.Logp(x[0], math.Exp(x[1]),
                      <- m.Data)
  }
  return ll
}
  \end{lstlisting}
  \caption{The basic example with the data read from a channel.}
  \label{fig:basic-example-channel}
\end{figure}

A similar approach can be applied to streaming data: the model
may read the data gradually as it arrives, and the inference
algorithm will emit samples from the time process inferred from
the data. Work is still underway on algorithms ensuring robust
and sound inference on streams, but Infergo models connected to
data sources via channels are already used in practice.

\section{Case Studies}
\label{sec:case-studies}

In this section, we present three case studies of probabilistic
programs, each addressing a different aspect of Infergo. Eight
schools (Section~\ref{sec:eight-schools}) is a popular example  we
use to compare model specification in Stan and Infergo. Linear
regression (Session~\ref{sec:linear-regression}) formulated as a
probabilistic program lets us explore a model comprising
multiple methods and illustrate the simulation-inference pattern.
Gaussian mixture (Section~\ref{sec:gaussian-mixture}) is
the most complex model of the three, featuring nested loops and
conditionals.

By no means do the models presented in this section reflect the
complexity or the size of real-world Infergo applications, the
latter reaching hundreds of lines of code in length and
performing inference on tens of thousands of data entries.
Rather, their purpose is to provide a general picture of
probabilistic programming with Infergo.

\subsection{Eight Schools}
\label{sec:eight-schools}

The eight schools problem~\cite{GCS+03} is an application of
the hierarchical normal model that often serves as an introductory
example of Bayesian statistical modeling. Without going into
details, in this problem the effect of coaching programs on test
outcomes is compared among eight schools.
Figure~\ref{fig:eight-schools} shows specifications of the same
model in Stan and Infergo, side-by-side.
\begin{figure*}
\begin{minipage}[t]{0.45\textwidth}
  \begin{lstlisting}[language=Stan,framexleftmargin=10pt,numbers=left]
data {
  int<lower=0> J;
  vector[J] y;
  vector<lower=0>[J] sigma;
}

parameters {
  real mu;
  real<lower=0> tau;
  vector[J] eta;
}

transformed parameters {
  vector[J] theta;
  theta = mu + tau * eta;
}

model {
  eta ~ normal(0, 1);
  y ~ normal(theta, sigma);
}
\end{lstlisting}

\centering
a. Stan
  \end{minipage}
  \hfill
\begin{minipage}[t]{0.5\textwidth}
\begin{lstlisting}[framexleftmargin=10pt,numbers=left]
type SchoolsModel struct {
  J          int
  Y          []float64
  Sigma      []float64
}


func (m *SchoolsModel) Observe(x []float64) float64 {
  mu := x[0]
  tau := math.Exp(x[1])
  eta := x[2:]

  ll := Normal.Logp(0, 1, eta)

  for i, y := range m.Y {
    theta := mu + tau*eta[i]
    ll += Normal.Logp(theta, m.Sigma[i], y)
  }

  return ll
}
\end{lstlisting}
\centering
b. Infergo
  \end{minipage}
  \caption{Eight schools: Stan vs. Infergo. The Go implementation
  has a similar length and structure to the Stan model.}
  \label{fig:eight-schools}
\end{figure*}

Stan models are written in a probabilistic programming language
tailored to statistical modeling. Infergo models are implemented
in Go. One can see that the models have roughly the same length
and structure. The definition of the model type in Go (lines
1--5) corresponds to the \lstinline{data} section in Stan (lines
1--5). Destructuring of the parameters of \lstinline{Observe} in
Go (lines 9--11) names the parameters just like the
\lstinline{parameters} section in Stan (lines 8--10).
Introduction of \lstinline{theta} in Go (line 16) corresponds to
the \lstinline{transformed parameters} section in Stan (lines
13--16). The only essential difference is the explicit loop over
the data in Go (lines 15--18) versus vectorized computations in
Stan (lines 15 and 20). However, vectorized computations are used
in Stan to speed-up execution more than to improve readability.
If repeated calls to \lstinline{Normal.Logp} (line 17) become
the bottleneck, an optimized vectorized version can be
implemented in Go in the same package. While subjective, our
feeling is that the Go version is as readable as the Stan one.
In addition, a Go programmer is less likely to resist adoption
of statistical modeling if the models are implemented in Go.

\subsection{Linear Regression}
\label{sec:linear-regression}

The model in Figure~\ref{fig:linear-regression} specifies linear
regression on unidimensional data: an improper uniform prior is
placed on the parameters $\alpha$, $\beta$, and $\log \sigma$
(lines 9--10), which are then conditioned on observations $x, y$
such that
\begin{equation}
y \sim \mathrm{Normal}(\alpha + \beta x, \sigma)
	\label{eqn:linear-regression-1}
\end{equation}
(lines 12--17).
\begin{figure}
\begin{lstlisting}[framexleftmargin=10pt,numbers=left]
type LRModel struct {
  Data  [][]float64
}

func (m *LRModel) Observe(x []float64)
    float64 {
  ll := 0.

  alpha, beta := x[0], x[1]
  sigma := math.Exp(x[2])

  for i := range m.Data {
    ll += Normal.Logp(
      m.Simulate(m.Data[i][0], alpha, beta),
	  sigma,
	  m.Data[i][1])
  }
  return ll
}

// Simulate simulates y based on x and 
// regression parameters.
func (m *LRModel) Simulate(x, alpha, beta float64)
    float64 {
  y := alpha + beta*x
  return y
}
\end{lstlisting}
\caption{Linear regression: the simulator is re-used in both
  inference and prediction.}
\label{fig:linear-regression}
\end{figure}
If only the maximum \textit{a posteriori} is estimated, the
parameter values will maximize their likelihood given the
observations:
\begin{equation}
	(\alpha, \beta, \sigma)_{MAP} = \arg \max \prod\limits_i \mathrm{pdf}_\mathrm{Normal}(\alpha + \beta x_i, \sigma)
	\label{eqn:linear-regression-2}
\end{equation}

Once the values of $\alpha$ and $\beta$ are estimated, they are
used to predict $y$ for unseen values of $x$. Hence, the linear
transformation $\alpha + \beta x$ is computed in two different
contexts: first, to compute the likelihood of the parameter
values; then, to predict $y$ based on a given $x$. This is an
example of the simulation-inference pattern
(Section~\ref{sec:challenges}): method \lstinline{Simulate}
simulates $y$ based on $x$ and is invoked both from
\lstinline{Observe} and then directly for prediction.  Since
both original and differentiated model are simultaneously
available in the source code, $Simulate$ can be called
for prediction on the original model, without any
differentiation overhead. 

In this greatly simplified case study, the simulator is a
one-line function. In a real-world scenario though, the
simulator can be laborious to re-implement correctly and
efficiently; for example, a simulator that executes a
parameterized policy in an uncertain or dynamic domain, and
depends on both parameter values and observations~\cite{MPT+16}.
The ability to re-use a single implementation both as a part of
the probabilistic program, where the simulator must be augmented
for inference (e.g. by automatic differentiation), and for
prediction or decision making, where the simulator may be
executed repeatedly and under time pressure, saves the effort
otherwise wasted on rewriting the simulator and removes the need
to maintain consistency between different implementations.

\subsection{Gaussian Mixture Model}
\label{sec:gaussian-mixture}

The Gaussian mixture model is a probabilistic model of a mixture
of components, where each component comes from a Gaussian
distribution. The model variant in
Figure~\ref{fig:gaussian-mixture} accepts a vector of
single-dimensional observations and the number of components,
and infers the distribution of parameters of each component
in the mixture. Following Stan User's Guide~\cite{SDT18},
component memberships are summed out to make the model
differentiable.
\begin{figure}
\begin{lstlisting}[framexleftmargin=10pt,numbers=left]
type GMModel struct {
  Data  []float64 // observations
  NComp int       // number of components
}

func (m *GMModel) Observe(x []float64) float64 {
  ll := 0.0
  // Impose a prior on component parameters
  ll += Normal.Logps(0, 1, x...)

  // Fetch the parameters
  mu := make([]float64, m.NComp)
  sigma := make([]float64, m.NComp)
  for j := 0; j != m.NComp; j++ {
    mu[j] = x[2*j]
    sigma[j] = math.Exp(x[2*j+1])
  }

  // Compute log likelihood of the mixture
  for i := 0; i != len(m.Data); i++ {
    var l float64
    for j := 0; j != m.NComp; j++ {
      lj := Normal.Logp(mu[j], sigma[j],
                        m.Data[i])
      if j == 0 {
        l = lj
      } else {
        l = logSumExp(l, lj)
      }
    }
    ll += l
  }
  return ll
}

// logSumExp computes log(exp(x)+exp(y)) robustly.
func LogSumExp(x, y float64) float64 {
  z := x
  if y > z {
    z = y
  }
  return z + math.Log(math.Exp(x-z)+math.Exp(y-z))
}

// logSumExp gradient must be supplied.
func init() {
  ad.RegisterElemental(logSumExp,
    func(_ float64, params ...float64)
      []float64 {
      z := math.Exp(params[1] - params[0])
      t := 1 / (1 + z)
      return []float64{t, t * z}
    })
}
\end{lstlisting}
\caption{Gaussian mixture model: an Infergo model with
  dynamic control structure and a user-defined elemental.}
\label{fig:gaussian-mixture}
\end{figure}
The model contains most of the steps commonly found in Infergo
models. First, a prior on the model parameters is imposed (lines
8--9). Then, the parameter vector is destructured into a form
convenient for formulation of conditioning (lines 11--17).
Finally, the parameters are conditioned on the observations
(lines 19--34). 

The conditioning of parameters involves a nested loop over all
observations and over all mixture components for each
observation.  Computing log-likelihood of the model parameters
implies summing up likelihoods over components for each
observation.  In the log-domain, a trick called `log-sum-exp' is
required to avoid loss of precision. The trick is implemented
as function \lstinline{logSumExp} and called for each
observation and component (line 28).

\lstinline{logSumExp} could have been implemented as a model
method and differentiated automatically. However, this function
is called on each iteration of the inner loop. Hence,
efficiency of inference in general is affected by 
an efficient implementation of \lstinline{logSumExp} (this can
also be confirmed by profiling). To save on the cost of a
differentiated call and speed up gradient computation,
\lstinline{logSumExp} is implemented as an elemental (lines
36--43). The hand-coded gradient is supplied for
\lstinline{logSumExp} (lines 45--54) to make automatic
differentiation work.

While still a very simple model compared to real-world Infergo
models, this case study illustrates common parts and coding
patterns of an elaborated model. The model code may be quite
complicated algorithmically and manipulate Go data structures
freely. When efficiency becomes an issue, parts of the model
can be implemented at a lower level of abstraction, but
nonetheless in the same programming language as the rest of the
model.

\section{Performance}

\begin{table*}
	\begin{tabular}{r | c |  c | c | c | c | c | }
		{\it model}         & \multicolumn{6}{c|}{\it time, seconds} \\
					        & \multicolumn{3}{c|}{\it compilation} & \multicolumn{3}{c|}{\it execution} \\
					        & Infergo & Turing & Stan              & Infergo  & Turing & Stan    \\\hline
              Eight schools & 0.50    & -      & 50                & 0.60     & 2.8    & 0.12    \\
     Gaussian mixture model & 0.50    & -      & 54                & 32       & 14     & 4.9     \\
Latent Dirichlet allocation & 0.50    & -      & 54                & 8.9      & 12     & 3.7     \\ 
	\end{tabular}
\caption{Compilation and execution times for 1000 iterations 
	of HMC with 10 leapfrog steps.}
\label{tab:memory-runtime}
\end{table*}
It is customary to include performance evaluation on benchmark
problems in publications on new probabilistic programming
frameworks~\cite{PW14,WVM14,TMY+16,GXG18}. We provide here a
comparison of running times of Infergo and two other probabilistic
programming frameworks: Turing~\cite{GXG18} and
Stan~\cite{Stan17}. Turing is a Julia~\cite{Julia14} library
and DSL offering exploratory statistical modeling  and
inference composition. Stan is a compiled language and a library
optimized for high performance.

We use three models of different size and structure for the
comparison: two of the models, Eight schools and Gaussian
mixture model, were introduced as a part of the case
studies~(Section~\ref{sec:case-studies}); the code for the third
model, Latent Dirichlet allocation~\cite{BNG03}, is included in
Appendix~\ref{app:lda}. The code and data for all models were
originally included in Stan examples and translated for use
with Turing and Infergo. \textit{Eight schools} is the smallest model.
One may expect that most of the time execution time is spent in
executing the boilerplate code rather than the model. \textit{Gaussian
mixture model} is a more elaborated model which benefits greatly
from vectorization. The greater the number of observations (1000
data points were used in the evaluation), the more significant
is the benefit of vectorized computations. \textit{Latent
Dirichlet allocation} is, again, a relatively elaborated model,
which is, however, not as easy to vectorize. 

We report both compilation time (for Stan and Infergo) and
execution time. Compilation time is an important characteristic
of a probabilistic programming framework --- statistical
modeling often involves many cycles of model modifications and
numerical experiments on subsampled data. The ability to modify
and instantly re-run a model is crucial for exploratory data
analysis and model design. Turing is based on Julia, which is
a dynamic language with JIT compiler, and there is no separate
compilation time. For Stan, the compilation time includes
compiling the Stan model to C++, and then compiling the C++
source into executable code for inference. For Infergo, the
compilation time consists of automatic differentiation of the
model, and then building a command-line executable that calls
an inference algorithm on the model.

Table~\ref{tab:memory-runtime} shows running time measurements
on the models.  The measurements were obtained on a 1.25GHz
Intel(R) Core(TM) i5 CPU with 8GB of memory for 1000 iterations
of Hamiltonian Monte Carlo with 10 leapfrog steps and averaged
over 10 runs.  In execution, Infergo is slower than Stan, which
comes as little surprise --- Stan is a mature highly optimized
inference and optimization library. However, Infergo and Turing
show similar execution times, despite Turing relying on Julia's
numerical libraries. The slowest relative execution time is on
the Gaussian mixture model, where Stan is 6--7 times faster,
apparently because Infergo model is not vectorized; the model
could be made more efficient at the cost of verbosity and poorer
readability, but we opted to use idiomatic Infergo models for
the comparison. On the Latent Dirichlet allocation model, run on
Stan example data with 25 documents and 262 word instances,
Infergo is only 2.5 times slower than Stan, and faster than
Turing, an illustration of the benefits of transformation-based
differentiation and efficiency of Go.

Infergo programs compile very fast, in particular in comparison
to Stan models. Both automatic differentiation and code
generation combined take a fraction of a second. Compilation time
of Stan models is dominated by compilation of the generated C++
code which takes close to a minute on our hardware. The ability
to modify a model and then re-run inference almost instantly,
and with reasonable performance, helps in exploratory model
development with Infergo, which was conceived with server-side,
unattended execution in mind, but fits the exploration and
development stage just as well.

\section{Conclusion}

Building a production software system which employs
probabilistic inference as an integral part of prediction and
decision making algorithms is challenging. To address the
challenges, we proposed guidelines for implementation of a
deployable probabilistic programming facility. Based on the
guidelines, we introduced Infergo, a probabilistic programming
facility in Go, as a reference implementation in accordance with
the guidelines.  Infergo allows to program inference models in
virtually unrestricted Go, and to perform inference on the model
using either supplied or third-party algorithms. We demonstrated
on case studies how Infergo overcomes the challenges by
following the guidelines.

A probabilistic programming facility similar to Infergo can be
added to most modern general-purpose programming languages, in
particular those used for implementation of large-scale software
systems, making probabilistic programming inference more
accessible in large scale server-side applications. We described
Infergo implementation choices and techniques in intent
to encourage broader adoption of probabilistic programming in
production software systems and ease implementation of similar
facilities in other languages and software environments.

Infergo is an ongoing project. Development of Infergo is
influenced both by feedback from the open-source Go community
and by the needs of the software system in which Infergo was
initially deployed, and which still supplies the most demanding
use cases in terms of flexibility, scalability, and ease of
integration and maintenance. Deploying probabilistic programming
as a part of production software system has a mutual benefit of
both improving the core algorithms of the system and of helping
to shape and improve probabilistic modeling and inference, thus
further advancing the field of probabilistic programming.

\end{sloppypar}

\clearpage
\clearpage

\bibliography{refs}

\clearpage

\appendix

\begin{figure*}
\section{Latent Dirichlet Allocation Model}
	\label{app:lda}
\vspace{\baselineskip}

\begin{minipage}[t]{0.49\textwidth}
\centering
\textbf{Infergo}

\begin{lstlisting}[framexleftmargin=10pt,numbers=left]
type LDAModel struct {
  K     int       // num topics
  V     int       // num words
  M     int       // num docs
  N     int       // total word instances
  Word  []int     // word n
  Doc   []int     // doc for word n
  Alpha []float64 // topic prior
  Beta  []float64 // word prior
}

func (m *LDAModel) Observe(x []float64) float64 {
  ll := 0.0

  // Regularize the parameter vector
  ll += Normal.Logps(0, 1, x...)
  // Destructure parameters
  theta := make([][]float64, m.M)
  m.FetchSimplices(&x, m.K, theta)
  phi := make([][]float64, m.K)
  m.FetchSimplices(&x, m.V, phi)

  // Impose priors
  ll += Dirichlet{m.K}.Logps(m.Alpha, theta...)
  ll += Dirichlet{m.V}.Logps(m.Beta, phi...)

  // Condition on observations
  gamma := make([]float64, m.K)
  for in := 0; in != m.N; in++ {
    for ik := 0; ik != m.K; ik++ {
      gamma[ik] = math.Log(theta[m.Doc[in]-1][ik]) +
        math.Log(phi[ik][m.Word[in]-1])
    }
    ll += D.LogSumExp(gamma)
  }
  return ll
}

func (m *Model) FetchSimplices(
	px *[]float64,
	k int,
	simplices [][]float64,
) {
	for i := range simplices {
		simplices[i] = make([]float64, k)
		D.SoftMax(model.Shift(px, k), simplices[i])
	}
}
\end{lstlisting}
\end{minipage}
\hfill
\begin{minipage}[t]{0.49\textwidth}
	\centering
	
	\textbf{Stan}

\begin{lstlisting}[language=Stan]
data {
  int<lower=2> K;               // num topics
  int<lower=2> V;               // num words
  int<lower=1> M;               // num docs
  int<lower=1> N;           // total word instances
  int<lower=1,upper=V> w[N];    // word n
  int<lower=1,upper=M> doc[N];  // doc for word n
  vector<lower=0>[K] alpha;     // topic prior
  vector<lower=0>[V] beta;      // word prior
}


parameters {
  simplex[K] theta[M]; // topic dist for doc m
  simplex[V] phi[K];   // word dist for topic k
}


model {
  for (m in 1:M)  
    theta[m] ~ dirichlet(alpha);  // prior
  for (k in 1:K)  
    phi[k] ~ dirichlet(beta);     // prior

  for (n in 1:N) {
    real gamma[K];
    for (k in 1:K) 
      gamma[k] <- log(theta[doc[n],k]) 
	  	+ log(phi[k,w[n]]);
    increment_log_prob(log_sum_exp(gamma));
  }
}
\end{lstlisting}
\end{minipage}
\end{figure*}

\end{document}